\documentclass[letterpaper, 10 pt, conference]{ieeeconf} \IEEEoverridecommandlockouts                              
\usepackage{mathptmx} 
\usepackage{times} 
\usepackage[libertine]{newtxmath}

\allowdisplaybreaks
\usepackage{float}
\usepackage{xifthen}
\usepackage{physics}
\usepackage{hyperref}
\usepackage{cleveref}
\usepackage{amsmath}
\usepackage{bbm}
\usepackage{thmtools}
\usepackage[dvipsnames]{xcolor}
\usepackage{comment}
\usepackage{graphicx}
\usepackage{parskip}
\usepackage{bm}
\usepackage{wrapfig}

\usepackage{makecell}

\let\originalleft\left
\let\originalright\right
\renewcommand{\left}{\mathopen{}\mathclose\bgroup\originalleft}
\renewcommand{\right}{\aftergroup\egroup\originalright}

\newcommand{\bP}[2][]{\Pr\ifthenelse{\isempty{#1}}{}{_{#1}}\left[#2\right]}
\newcommand{\bE}[2][]{\mathop\mathbb{E}\ifthenelse{\isempty{#1}}{}{_{#1}}\left[#2\right]}
\newcommand{\bI}[2][]{\mathop\mathbb{I}\ifthenelse{\isempty{#1}}{}{_{#1}}\left[#2\right]}
\newcommand{\Var}[2][]{\mathbf{Var}\ifthenelse{\isempty{#1}}{}{_{#1}}\left[#2\right]}

\newcommand{\ds}[1]{\textcolor{black}{#1}}

\newcommand{\R}{\mathbb{R}}

\newcommand{\cK}{\mathcal{K}}
\newcommand{\cP}{\mathcal{P}}
\newcommand{\cQ}{\mathcal{Q}}

\newcommand{\F}{\mathcal{F}}
\newcommand{\G}{\mathcal{G}}

\title{\LARGE \bf 
Machine Learning Accelerated PDE Backstepping Observers}

\author{Yuanyuan Shi$^{1}$, Zongyi Li$^{2}$, Huan Yu$^{3}$, Drew Steeves$^{4}$, Anima Anandkumar$^{2}$, and Miroslav Krstic$^{4}$
\thanks{$^{1}$Yuanyuan Shi is with the Department of Electrical and Computer Engineering, University of California San Diego.
{\small yyshi@eng.ucsd.edu}}%
\thanks{$^{2}$Zongyi Li and Anima Anandkumar are with the Department of Computing and Mathematical Sciences, Caltech. {\small zongyili@caltech.edu}}%
\thanks{$^{3}$Huan Yu is with the Hong Kong University of Science and Technology (Guangzhou), Thrust of Intelligent Transportation, Guangdong, China.} 
\thanks{$^{4}$Drew Steeves and Miroslav Krstic are with the Department of Mechanical and Aerospace Engineering, University of California San Diego.}%
}

\begin{document}

\maketitle
\thispagestyle{empty}
\pagestyle{empty}

\begin{abstract}

State estimation is important for a variety of tasks,
from forecasting to substituting for unmeasured states in feedback controllers. Performing \ds{real-time} state estimation for PDEs using provably and rapidly converging observers, such as those based on PDE backstepping, is computationally expensive and in many cases prohibitive. 
We propose a framework for accelerating PDE observer computations using learning-based approaches that are much faster while maintaining accuracy. 
In particular, we employ the recently-developed Fourier Neural Operator (FNO) to learn the functional mapping from the initial observer state and boundary measurements to the state estimate.
\ds{By} employing backstepping observer gains for \ds{previously-designed} observers \ds{with particular convergence rate guarantees,}
we provide numerical experiments that evaluate the \ds{increased} computational efficiency gained with FNO. We consider the state estimation for three benchmark PDE examples motivated by applications: first, for a reaction-diffusion (parabolic) PDE whose state is estimated with an exponential rate of convergence; second, for a parabolic PDE with \ds{exact} prescribed-time estimation;
and, third, for a pair of coupled first-order hyperbolic PDEs \ds{that} modeling traffic flow density and velocity. 
The ML-accelerated observers trained on simulation data sets for these PDEs achieves up to three orders of magnitude improvement in computational speed compared to classical methods. This demonstrates the attractiveness of the ML-accelerated observers for real-time state estimation and control.
\end{abstract}

\section{INTRODUCTION}

Historically, the study of observability of PDEs goes back to at least the 1960s, whereas infinite-dimensional Kalman filters emerge \ds{later}
in the early 1970s. However, the associated operator Riccati equations in \ds{these} optimal state estimators 
are nonlinear infinite-dimensional equations, exacting a high price in not only numerical computation but also in the numerical analysis skills demanded from the user of such a state estimator. To address these challenges, a {\em PDE backstepping} approach for observer design was introduced in the 2005 paper 
~\cite{smyshlyaev2005backstepping}. Backstepping provides three properties that match or exceed the properties of other PDE observer design methods: (1) convergence to the state of the actual PDE system is guaranteed; (2) convergence rate can be assigned (made arbitrarily rapid) using the design parameters in the backstepping approach; (3) the gain functions in the backstepping approach can be explicitly computed with arbitrary accuracy because these gain functions satisfy linear integral equations of Volterra's second kind, rather than being governed by operator Riccati equations or nonlinear PDEs. 

The PDE backstepping observers~\cite{smyshlyaev2005backstepping} are of what is commonly referred to as the ``Luenberger structure,'' consisting of a copy of the PDE model and an output estimation error term, multiplied by a gain function that depends on the spatial variable(s). The PDE backstepping design provides the gain function, without the need to solve any PDE in real time. However, the observer itself is a PDE whose input is the measured output of the original control system. Conducting state estimation in real time, in a fashion that requires a rapid solution to a PDE, may be challenging. But rapid estimation is necessary for many applications, including those where the state estimate from the PDE observer is used in a state feedback controller for stabilization, as in~\cite{smyshlyaev2005backstepping}. 

It is for this reason that we are interested in methods that eliminate the need for numerically solving the observer PDEs in real time. We pursue this in the present paper using an FNO approximation of the observer PDEs. 

{\bf Literature Overview on PDE Backstepping Observers.}\footnote{Even though this literature review may appear as an indulgence in self-citation, the objective is to provide the reader with helpful information on the nearly 20-year history of the subject of PDE backstepping observers.} Following the introduction of the PDE backstepping observer design approach in~\cite{smyshlyaev2005backstepping} for parabolic 1-D PDEs, numerous extensions of this approach to hyperbolic and other PDE structures have followed. Extensions to wave PDEs were introduced in~\cite{KRSTIC200863}, and to beam models arising in atomic force microscopy in~\cite{doi:10.1137/060676969}. 
An adaptive observer for a single first-order hyperbolic PDE is reported in~\cite{BERNARD20142692}. Designs for coupled systems of first-order hyperbolic PDEs are given in~\cite{6573344}, including an experimental application to off-shore oil drilling in~\cite{HASAN201675}. 
Generalizations from observers for PDEs to backstepping observers for ODE-PDE-ODE ``sandwich'' systems are given in~\cite{8651518} for the parabolic class and in \cite{9151182} for the hyperbolic class. Observer extensions to less standard PDE classes are given in \cite{doi:10.1137/070704290} for the Schr\"{o}dinger PDE, in \cite{4154955} for the Ginzburg-Landau PDE model of vortex shedding, and in \cite{10.1115/1.3023128} for the nonlinear Burgers PDE. 
Observers for PDE systems formulated in higher dimensions in polar and spherical coordinates are given for the annulus geometry in \cite{5208259} for thermal convection and in \cite{refId0} for PDEs on arbitrary-dimensional balls. All these results ensure exponential convergence of the state estimate. A design for ``prescribed-time'' observer convergence (independent of the distance between the system initial condition and the estimate initial condition) is given in \cite{8796114}.
In addition to the applications mentioned above~\cite{doi:10.1137/060676969,HASAN201675,10.1115/1.3023128}, backstepping PDE observers for the following applications have also been developed: for Lithium-ion batteries in \cite{7489035}, 
for the Stefan PDE-ODE model of phase transition arising in additive manufacturing~\cite{8358218}, 
and Arctic sea ice estimation \cite{KOGA2020108713}, 
for the Navier-Stokes and magnetohydrodynamic models of turbulent fluid flows in \cite{VAZQUEZ20082517}, and for traffic flows in \cite{yu2020pde}. 


{\bf Contributions.}
We use the Fourier neural operator as a surrogate model to accelerate the PDE backstepping observers. We propose two neural observer formulations: feedforward and recurrent. 
The former  is similar to the smoothing process: the neural observer is given the observations on a time interval to estimate the solution on the interval. The latter  is similar to the smoothing filtering process: the neural observer is given the observation at one time step and to predict for one time step. The feedforward neural observer gets higher accuracy whereas the recurrent neural observer holds more promise for high-frequency sample-data control. 

The paper's numerical results are given in Section \ref{sec-experiments}, where three examples of accelerated PDE backstepping observer implementations are presented: exponential and prescribed-time observers for certain 1-D reaction-diffusion PDE examples, followed by a coupled first-order hyperbolic PDE model of stop-and-go oscillations in congested traffic. 
In all examples, the FNO observers, once trained \ds{over a few hours} using a large set of initial conditions,
 perform real-time calculations up to 1000x faster than the basic PDE solvers. 

{\em No theorems are stated in the paper since, for the three experimental results in Section \ref{sec-experiments}, the observer convergence theorems are provided, respectively, in \cite[Theorem 8]{smyshlyaev2005backstepping}, \cite[Proposition 5.3]{8796114}, \cite[Theorem 3]{yu2020pde}.}




\section{Neural Operator}
\label{sec-FNO}


When \ds{implementing}
backstepping observers, \ds{numerical solvers}
such as finite difference methods and finite element methods \ds{are typically used} to solve the boundary value problems. These conventional numerical solvers \ds{may be} subjective to stability constraints \ds{(depending on the class of PDE)} and therefore \ds{smaller time steps need to be used,}
which is computationally expensive. 
Recently, machine learning (ML) methods have shown promise in solving PDEs.
Examples include impressive speedups in complex systems \ds{that model systems governing}
weather forecasting \cite{pathak2022fourcastnet} and \ds{producing numerical solutions to}
previously unsolvable problems such as high-dimensional PDEs \cite{han2018solving}. Inspired by these recent advances, we propose to use the data-driven neural operator as a surrogate solver for solving backstepping PDE observers. The neural operator solver shows orders of magnitude of speedup, making it \ds{more amenable to real-time applications}.

ML methods for solving PDE frame the problem as a statistical inference task. In the standard \ds{supervised}
learning task, a dataset \ds{containing}
inputs (boundary conditions or initial conditions) and outputs (solution functions) are given, and the task is to approximate the solution operator between \ds{the} input and output function spaces. Once such solution operator is learned, it can directly evaluate for any new input queries without solving any \ds{PDEs.}
As a result, the data-driven solver can be orders of magnitude faster compared to conventional solvers.

The model design is one of the most critical questions in ML. An ideal model \ds{has}
a proper structure that takes fewer data points to train and generalizes better to new queries. 
\ds{In~\cite{lu2019deeponet, li2021fourier},} neural operators are designed for solving partial differential equations and dynamical systems.
They consist of integral operators and nonlinear activation functions so that their representation capability is general enough to approximate complex nonlinear operators. \begin{figure}[t]
    \centering
    \includegraphics[width=\linewidth]{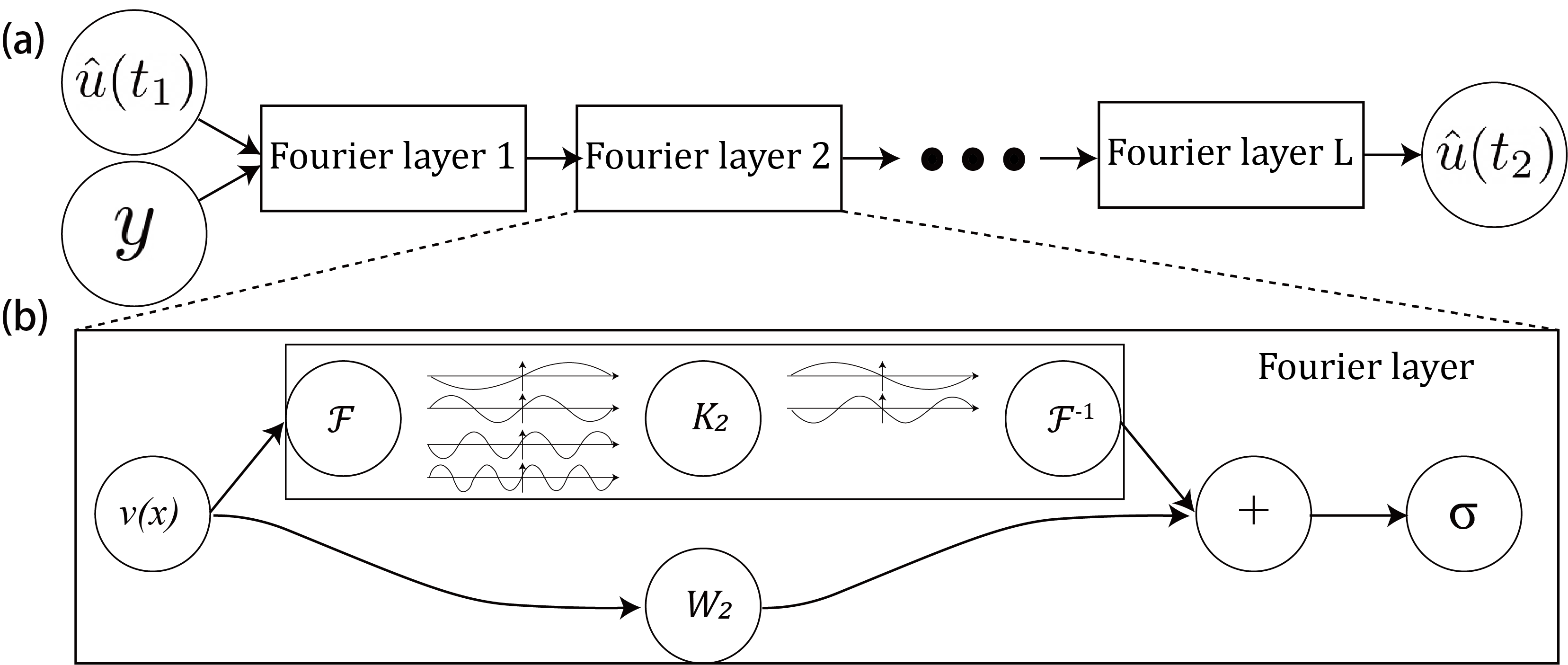}
    \caption{Neural Observer with FNO backbone (Figure is adapted from \cite{li2021fourier} 
    }
    \label{fig:FNO}
\end{figure}
As shown in Figure \ref{fig:FNO}, \ds{the} neural operator \cite{li2021fourier} composes linear integral operator $\cK$ with pointwise non-linear activation function \(\sigma\) to approximate highly non-linear operators.
\begin{equation}
\label{eq:G}
    \G_{\theta} \coloneqq \cQ \circ\sigma(W_{L} + \cK_{L}) \circ \cdots \circ \sigma(W_1 + \cK_1) \circ \cP
\end{equation}
where \(\cP: \R^{d_a} \to \R^{d_{1}}\), \(\cQ: \R^{d_{L}} \to \R^{d_u}\) are the pointwise neural networks that encode the lower dimension function into higher dimensional space and vice versa. The model stack $L$ layers of $\sigma(W_{l} + \cK_{l})$ where \(W_l \in \R^{d_{{l+1}} \times d_{l}}\) 
are pointwise linear operators (matrices), 
\(\cK_l: \{D \to \R^{d_{l}}\} \to \{D \to \R^{d_{l+1}}\}\) are integral kernel operators. We use ReLU activation in all numerical experiments, where $\sigma(x) = max(0, x)$ element-wise. 
The parameters $\theta$ of the neural operator to be learned consists of all the parameters in $\cP, \cQ, \{W_l, \cK_l\}_{l=1}^{L}$.
%


In the Fourier neural operator (FNO)~\cite{li2021fourier}, the integral kernel operators are restricted to convolution, and implemented with Fourier transform:
\begin{equation}
\label{eq:Fourier}
\bigl(\cK v_t\bigr)(x)=   
\F^{-1}\Bigl(K \cdot (\F v_t) \Bigr)(x) \qquad \forall x \in D .
\end{equation}
where $\F$ denotes the Fourier transform of a function $f: D \rightarrow \mathbb{R}^{d_v}$, $\F^{-1}$ 
denotes its inverse, and $K$ denotes the linear transformation matrix applied to function $(\F v_t)$ in Fourier domain. An illustration is given in Figure~\ref{fig:FNO}.

Leveraging FNO's powerful representation ability and superior computation efficiency, we propose to use it for learning and accelerating PDE observer computations. As we show here through our experiment results, FNO achieves up to \emph{three orders of magnitude} improvement in computational speed compared to traditional PDE solvers, making FNO-PDE observers suitable for real-time control \ds{and monitoring}. 

\section{METHODS}
\label{sec-methods}
In this section, we describe the proposed ML-accelerated PDE backstepping observers. We start with introducing the structure of a Learning-type PDE observer; \ds{we then}
present two model structures for implementation. 

\subsection{A PDE observer with boundary sensing}

We consider a general nonlinear PDE system on a 1-D spatial interval $[0,1]$,
\begin{equation}\label{eq:general_dyn}
u_t = f_2(u)u_{xx}+f_1(u) u_x +f_0(u)\,,
\end{equation}
where $u(x, t)$ denotes the system state, whose arguments will be suppressed wherever possible for the sake of legibility. $f_i$ are real-valued functions of the \ds{state,} and assume $f_i$ is differentiable over the state space.
Suppose the boundary value $y(t)=u(0,t)$ is available for measurement.

We formulate a Luenberger-type PDE observer 
\begin{equation}\label{eq:general_obs}
\hat u_t = f_2(\hat u)\hat u_{xx}+f_1(\hat u) \hat u_x +f_0(\hat u)+p(x)\left[y(t)-\hat u(0,t)\right]\,,
\end{equation}
where the first three terms are a copy of the system 
\eqref{eq:general_dyn} and $p$ is a gain function of the observer, to be designed (typically using the PDE backstepping approach). Denoting the state estimation error $\tilde u = u - \hat u$ and subtracting \eqref{eq:general_obs} from \eqref{eq:general_dyn}, we get the observer error system
\begin{align}\label{eq-obserrsys}
    \tilde u_t = & \psi_2(u,\tilde u) \tilde u_{xx}+\psi_1(u,\tilde u) \tilde u_{x}\nonumber\\
    &+\left[\phi_2(u,\tilde u) u_{xx}+\phi_1(u,\tilde u) u_x +\phi_0(u,\tilde u) \right]\tilde u\nonumber\\
    &+p(x)\tilde u(0,t)\,,
\end{align}
where $\psi_i(u,\tilde u)= f_i(u) - \phi_i(u,\tilde u)\tilde u$ and $\phi_i$ is defined, using the mean value theorem, through $f_i(a)- f_i(b) = \phi_i(a,a-b)(a-b)$.

The goal of observer design is for the estimation error $\tilde u(x,t) = u(x,t) - \hat u(x,t)$ to 
\ds{converge} to zero as time goes to infinity, \ds{with respect to} 
a particular spatial norm.
This is to be achieved by making the origin of the PDE \eqref{eq-obserrsys} exponentially stable through the choice of the observer gain function $p(x)$. Using the PDE backstepping approach \cite{smyshlyaev2005backstepping}, such a stabilizing $p(x)$ can be found, under suitable conditions on the functions $f_i$ (e.g., for constant $f_2, f_1$ and linear $f_0$, but also much more generally) and on the \ds{regularity of the} observer \ds{initial error}.


To obtain the state estimate $\hat u(x,t)$ in real time, one needs to \ds{simulate}
the observer \eqref{eq:general_obs} in real time. This observer is a PDE itself, with $y(t)$ as its input. The PDE  \eqref{eq:general_obs} does not have an explicit solution in general, and \ds{numerically generating its solution}
may be prohibitively slow for certain real-time applications such as \ds{for} feedback \ds{control}. 


\subsection{ML-accelerated PDE backstepping observers}
Motivated by the challenges \ds{highlighted} above, the question we address in this paper is: 

\emph{Can we use machine learning to design computationally-efficient PDE observers while maintaining rigorous performance guarantees in state estimation?}

We answer the question affirmatively by designing an ML-accelerated PDE observer framework that learns a functional mapping from the initial observer state and system boundary measurements, to the state estimation. To guarantee the convergence of the neural observer to the state of the actual PDE system, we generate supervised training data from PDE observers designed by \ds{the} {\em backstepping} method. To accelerate the computational efficiency, we use FNO for model learning, which shows superior performance in learning mappings between infinite-dimensional spaces of functions.
Pairing the PDE backstepping design of observer gains with an FNO approximation of the observer PDEs resulting from backstepping inherits the mathematical rigor of the former (backstepping) and the computational performance of the latter (FNO)\ds{---allowing}
us to achieve the best of both worlds. 

Specially, we propose two types of ML-accelerated PDE observers, i) \ds{the} Feedforward Neural Observer, which learns the mapping from the initial state and the entire sequence of system boundary measurements, to the state estimation; and ii) \ds{the} Recurrent Neural Observer, which learns a mapping from the previous
step's state estimation and boundary measurement to the next time step's state estimation. The learned mapping is then applied recursively over the temporal domain to obtain the overall state estimation.


\subsubsection{Feedforward Neural Observer} 
\begin{figure}[t]
  \centering
  \framebox{\parbox{3in}{\includegraphics[width=3in]{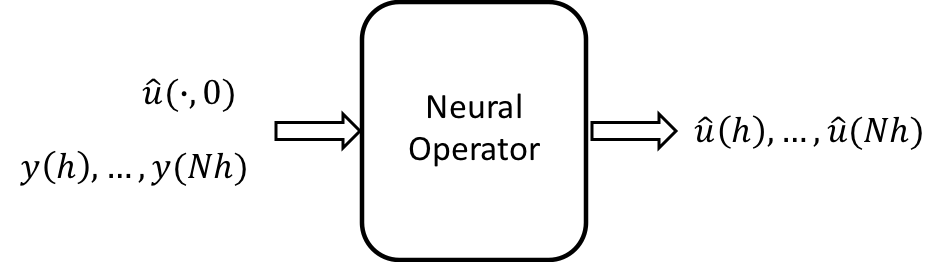}}}
  \caption{Diagram of the Feedforward Neural Observer.}
  \label{fig:approach1}
\end{figure}

Consider a time horizon $[0, T]$ with a measurement interval $h$ and contains $N$ intervals: $t = h, 2h, \cdots, Nh$. Here and hereafter, we use $T, h, N$ to denote the entire system operation horizon, the time interval between two measurements, and the total number of measurements.
For the feedforward neural observer design, we learn a neural operator
$\G_{\theta}$
that directly maps the initial \ds{estimate}
$\hat{u}(\cdot, 0)$ and system measurements $\{y(h), \dots, y(Nh)\}$ to \ds{the estimates}
$\{\hat{u}(\cdot, h),\dots,\hat{u}(\cdot, Nh)\}$. 
Figure \ref{fig:approach1} \ds{illustrates}
the model structure of this direct prediction approach.

\subsubsection{Recurrent Neural Observer}
\begin{figure}[t]
  \centering
  \framebox{\parbox{3in}{\includegraphics[width=3in]{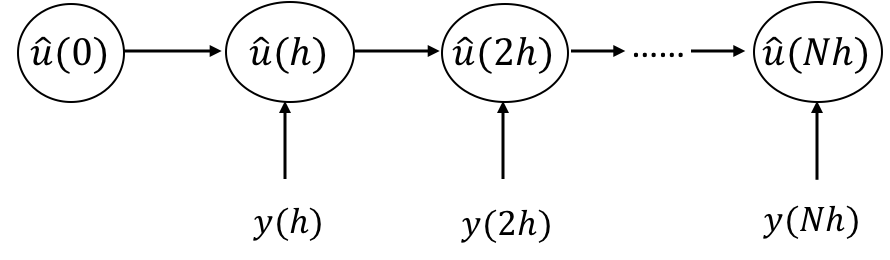}}}
  \caption{Diagram of the Recurrent Neural Observer}
  \label{fig:approach2}
\end{figure}

Unlike the first approach where inputs over time $[0, T]$ are fed into the neural operator as a full vector, the recurrent approach feeds input sequentially. At time step $t$, it uses the previous step state estimate $\hat{u}(\cdot, (t-1)h)$, and the input $y((t-1)h))$ to obtain the current timestep state estimate $\hat{u}(\cdot, th)$ via the following computation:
$$\hat{u}(\cdot, th) = \G_{\theta}\left(\hat{u}(\cdot, (t-1)h), y((t-1)h)\right)$$
where $\G_{\theta}$ is the neural operator. \ds{The state estimate} $\hat{u}(\cdot, th)$ is passed into the neural operator at time $(t+1)h$ with the same parameters. Figure \ref{fig:approach2} shows
the model structure of the recurrent neural observer.

Comparing the two model structures, we observed that the feedforward neural observers tend to obtain better estimation accuracy and train faster compared to their recurrent-structured counterparts in simulations, which is suitable for smoothing process where the neural observer is given the observations on a time
interval to estimate the solution on the interval. On the other hand, the recurrent neural observer produces system estimation at each timestep, it can be applied to applications of observer-based controllers. 



\section{EXPERIMENTS}
\label{sec-experiments}

To illustrate the capability of the proposed ML-accelerated PDE observers, we compare the proposed observers with the conventional PDE observer on a chemical tubular reactor example (nonlinear parabolic PDEs) with both an exponential convergent backstepping observer and a prescribed-time convergent backstepping observer, and a backstepping traffic observer (nonlinear hyperbolic PDEs). 

In order to demonstrate the effectiveness of FNO-backstepping pairing, we add a baseline long short-term memory (LSTM) model, which is a commonly used deep learning model capable of learning dependence in sequence prediction. 
Unless otherwise specified, we use $1000$ training instances and 100 testing instances. We use Adam optimizer
to train for 500 epochs\footnote{Number of epochs defines the number times that the learning algorithm will work through the entire training dataset.} with an initial learning rate of $0.001$ that is halved every 100 epochs. Both the recurrent neural operator and LSTM are trained with the teacher-forcing \cite{lamb2016professor} that takes ground truth values as supervision during training.



\subsection{Exponentially Convergent Observer for a Reaction-Diffusion PDE}
We start with a chemical tubular reactor system, the dynamics of which are represented as,
\begin{subequations}\label{dyn:chem}
\begin{align}
    u_t(x, t) &= \epsilon u_{xx}(x, t) + \lambda_{\alpha\beta}(x) u{(x, t)} \,, \label{eq:chem1}\\
    u(0, t) &= 0 \,, \quad u_x(1, t) = 0 \label{eq:chem2}
 \end{align}
\end{subequations}
where $u$ is the temperature, $x \in (0, 1)$, and $t \in (0, T]$.
Equations \eqref{eq:chem1} and \eqref{eq:chem2} describe the heat transfer process within a heat generation chemical reaction~\cite{smyshlyaev2005backstepping}. 
The coefficient function $\lambda_{\alpha, \beta}$ is a ``one-peak'' function, $\lambda_{\alpha, \beta}(x) = \frac{2 \epsilon \alpha^2}{\cosh^2(\alpha x-\beta)}$ where the peak location and value are controlled by $\alpha$ and $\beta$. 
We use $T = 0.125$ second, $\alpha= 4, \beta = 2, \epsilon = 1$ in the experiments. 
The open-loop system is unstable with $u(1) = 0$, which makes the estimation problem challenging.

A backstepping observer for system \eqref{dyn:chem} is given by~\cite{smyshlyaev2005backstepping}
\begin{subequations}\label{eq:obs_reactor}
\begin{align}
    \hat{u}_t(x, t) &= \epsilon \hat{u}_{xx}(x, t) + \lambda_{\alpha\beta}(x) \hat{u}{(x, t)} + p_1(x)[u(1)-\hat{u}(1)]\,, \\
    \hat{u}(0, t) &= 0 \,, \quad \hat{u}_x(1, t) = 0
 \end{align}
\end{subequations}
with the backstepping gain function
\begin{align}\label{eq:obs_gain_reactor}
p_1(x) = \epsilon \alpha^2 \tanh(\beta)  (\tanh \beta - \tanh (\beta-\alpha x)){\rm e}^{(1-x) \alpha \tanh \beta} \,.
\end{align}
The exponential convergence of this observer is guaranteed by \cite[Theorem 8]{smyshlyaev2005backstepping}.


To measure the performance of the model, we use the relative L2 error defined as follows,
$$\mathcal{E} =\frac{1}{N} \sum_{i=1}^{N_{test}} \frac{\|\hat{u}^{(i)} - u^{(i)}\|_2}{\|u^{(i)}\|_2}\,,$$
where $u^{(i)}$ is the state estimation from the backstepping PDE observer in Eq~\eqref{eq:obs_reactor} and \eqref{eq:obs_gain_reactor} for test instance $i$, and $\hat{u}^{(i)}$ is the state estimation from the learning based observers. 

Table \ref{table_example1} shows that training and test error of three ML-accelerated observers. 
Both neural operators perform significantly better than the LSTM model. 
We further compare the computation efficiency of the ML-accelerated observers against the backstepping PDE observers. As evident from Table \ref{table_stats}, 
the ML-accelerated observers achieve significant speedup compared to the classic finite-difference PDE solver. In particular, the feedforward neural observer can shorten the computation time by more than three orders of magnitude, while maintaining accurate state estimation. 
\begin{table}[t]
\caption{Experiment Results on Exponential Convergent Observer for Chemical Tubular Reactor}
\label{table_example1}
\begin{center}
\begin{tabular}{l|l |l | l}
\hline
\bf{Model} & \makecell{\bf{Feedforward} \\ \bf{Neural Observer}}  & \makecell{\bf{Recurrent} \\ \bf{Neural Observer}} & \bf{LSTM} \\
\hline
Parameters & 2,368,033 & 549,633 & 4,917,101\\
\makecell{Total Training \\ Time (Hours)}  & 0.37 & 4.45 &0.069\\
Training Error & 7.28e-4 & 2.95e-3 &0.466\\ 
Test Error&  5.68e-4 & 0.1903 & 0.595\\ 
\hline
\end{tabular}
\end{center}
\end{table}

In addition to the significant acceleration of the neural observers compared to conventional PDE observers, we also hope to point out that the ``price'' is being paid in terms of \emph{upfront computation cost}. On the one hand, to train the neural observers, we need to generate training data via conventional solvers. For example, we collected 1000 training data with a finite-difference solver given different initial conditions, which accounts for $\sim 2$h upfront computation time. In addition, the feedforward neural operator training takes 500 epochs and each epoch takes 2.64s which needs $0.37$h total training time. For the recurrent neural operator, the per epoch training time is 32.05s and the total training time is $4.45$h. 

\begin{table}[t]
	\renewcommand{\arraystretch}{1.5}
	\centering
	\caption{Computation Time Comparison between Conventional PDE Observer and Various ML-Accelerated Observers for Reaction-Diffusion PDE.}
	\begin{tabular}{ccc}
		\hline
		\hline
        \bf{Method} & \bf{Time (seconds/instance)}\\
		\hline
		Conventional PDE Observer & 7.76 \\
		Feedforward Neural Observer & 0.0065\\
		Recurrent Neural Observer & 0.188 \\
		LSTM & 0.0237\\
		\hline
		\hline
	\end{tabular}
	\label{table_stats}
\end{table}

\subsection{Prescribed-Time  Observer for a Reaction-Diffusion PDE}
In addition to the exponentially convergent observers, we demonstrate the performance of the ML-accelerated {\em prescribed-time} (PT) convergent observers. For the same chemical tubular reactor system in \eqref{dyn:chem} with time-invariant coefficient $\lambda$, following \cite[Proposition 5.3]{8796114}, the backstepping PDE observer is designed as,
\begin{equation}    
\label{eq-PTobs}
    \hat u_t(x, t) =  \hat u_{xx}(x, t) +\lambda \hat u(x, t)+ p(x,t)\left[u_x(1,t) -\hat u_x(1,t) \right]\,,
\end{equation}
which employs the boundary measurement $u_x(1,t)$ and the time-varying backstepping gain function 
\begin{align} \label{eq-p(x,t)}
    p(x,t) &=  \mu T^3 {x\over 2 (T-t)^3}\sum_{n=0}^\infty \left({1-x^2\over 4(T-t)} \right)^n {(-1)^n\over (n+1)!} \\
    & \quad \times \sum_{j=0}^n {1\over j!} \left({-\mu T^3\over 2(T-t)^2} \right)^j\sum_{k=0}^j {j \choose k}{n+2+k \choose n-j}\,, \quad \mu>0\,,
\end{align}
guarantees the convergence of $\hat u(x,t)$ to $u(x,t)$ by $t=T$, for arbitrarily short $T>0$. This poses stringent requirements for the accuracy of neural observers since they're required to obtain same prescribed-time convergence behavior. 

\begin{figure}[t]
    \centering
    \begin{minipage}[t]{0.24\textwidth}
        \centering
        \includegraphics[width=1\linewidth]{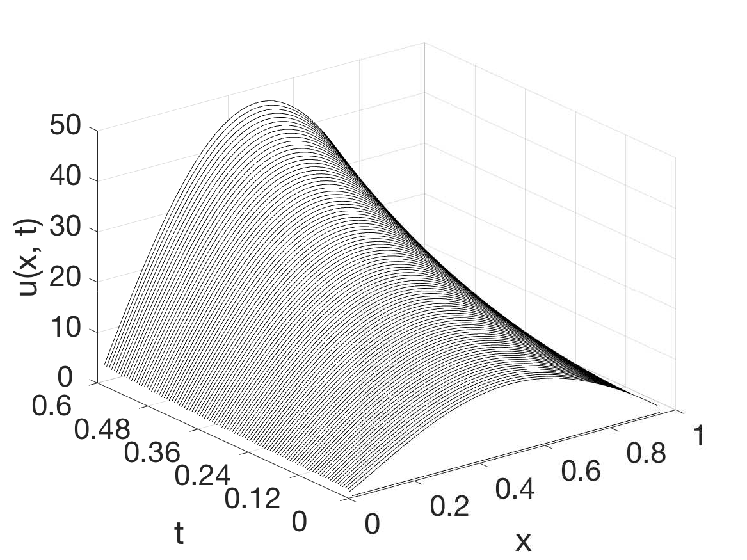}
    \end{minipage}\hfill
    \begin{minipage}[t]{0.24\textwidth}
        \centering
        \includegraphics[width=1\linewidth]{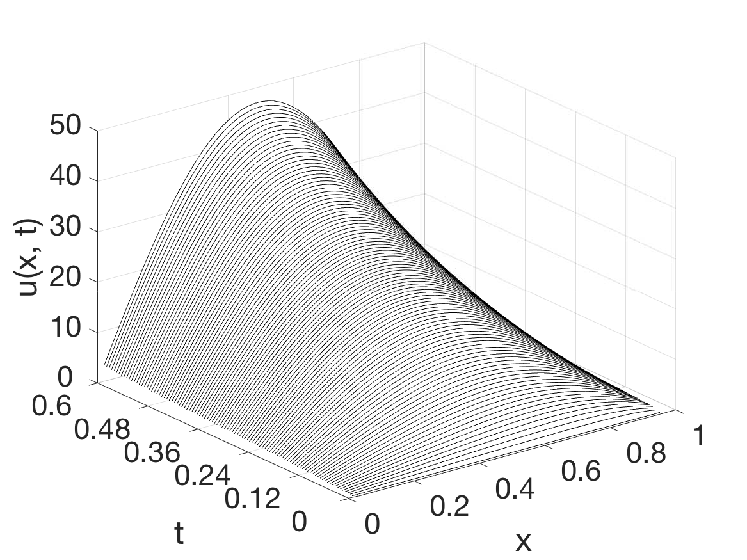}
    \end{minipage}\hfill
    \begin{minipage}[t]{0.24\textwidth}
        \centering
        \includegraphics[width=1\linewidth]{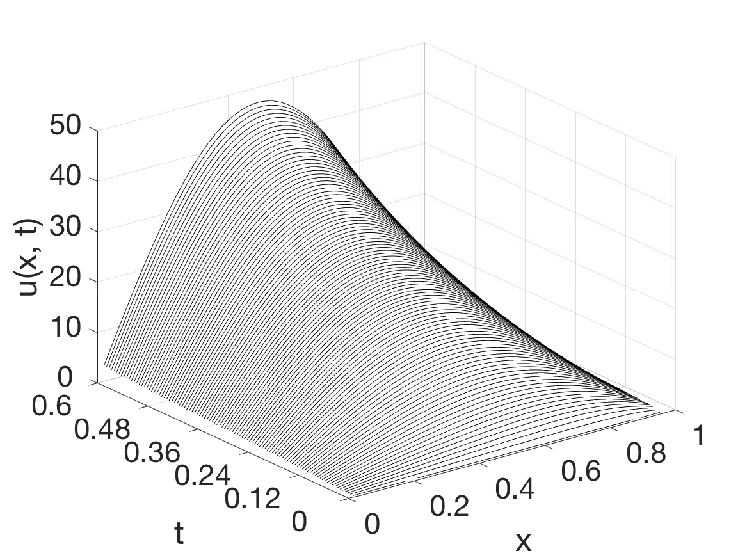}
    \end{minipage}\hfill
    \begin{minipage}[t]{0.23\textwidth}
        \centering
        \includegraphics[width=1\linewidth]{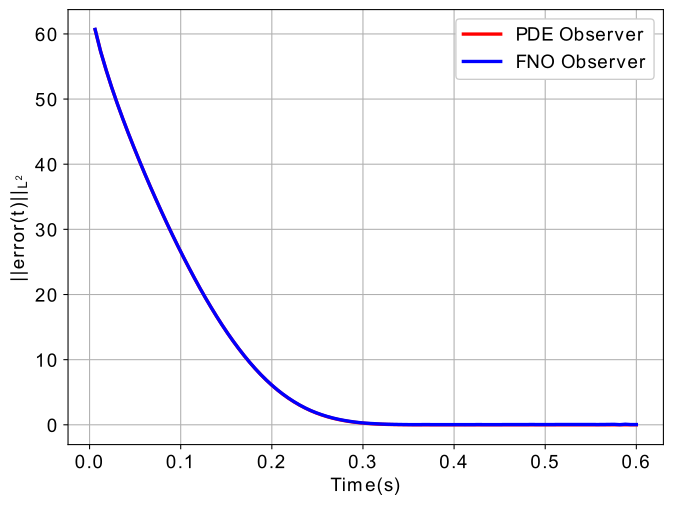}
    \end{minipage}\hfill
    \caption{Comparison of the true PDE system dynamics (top left), the estimation with a backstepping PDE observer with conventional solver (top right), and the feedforward neural observer (bottom left), under a unseen test initial condition. Bottom right figure shows the evolution of the estimation error of the backstepping PDE observer and FNO observer. }
\label{fig:example_PTobserver}
\end{figure}

In the experiment, the reaction–diffusion equation \eqref{eq:obs_reactor} is defined with parameters $\epsilon = 1$ and $\lambda_{ab} = 12$ (constant), which is unstable without control input. The prescribed convergent time defined as $T = 0.6$ second. For the backstepping PDE observer, we solve via an implicit Euler method with discretization $\Delta x = 0.02$ and $\Delta t = 0.006$, and we approximate \eqref{eq-p(x,t)} by limiting its characterization to the first eight terms of their infinite series. Figure \ref{fig:example_PTobserver} shows that the true system dynamics, the PDE backstepping observer \eqref{eq-PTobs} and the feedforward neural observer for one test case. Table \ref{table_example2} compares the performance of different ML-accelerated observers and Table \ref{table_example2_time} compares the computational efficiency. The acceleration with the feedforward neural observer is, again, about a thousandfold.



\begin{table}[t]
\caption{Experiment Results on Prescribed-Time Convergent Observer for Chemical Tubular Reactor}
\label{table_example2}
\begin{center}
\begin{tabular}{l|l |l | l}
\hline
\bf{Model} & \makecell{\bf{Feedforward} \\ \bf{Neural Observer}}  & \makecell{\bf{Recurrent} \\ \bf{Neural Observer}} & \bf{LSTM}\\
\hline
Parameters & 2,368,033 & 549,633 & 4,440,048\\
\makecell{Total Training \\ Time (Hours)}  & 0.24 & 2.42 & 0.067\\
Training Error & 3.84e-4 & 2.51e-3  & 4.64e-4\\ 
Test Error &  4.58e-4 & 9.05e-2  & 0.609\\ 
\hline
\end{tabular}
\end{center}
\end{table}
\begin{table}[t]
	\renewcommand{\arraystretch}{1.5}
	\centering
	\caption{Computation Time Comparison between Conventional PDE Observer and ML-Accelerated Observers for Prescribed-Time Observers.}
	\begin{tabular}{ccc}
		\hline
		\hline
        \bf{Method} & \bf{Time (seconds/instance)}\\
		\hline
		Conventional PDE Observer & 11.308 \\
		Feedforward Neural Observer & 0.0058 \\
		\hline
		\hline
	\end{tabular}
	\label{table_example2_time}
\end{table}


\vspace{-6pt}
\subsection{Traffic Flow PDE Observer}
Finally, we demonstrate the performance of the ML-accelerated observers on 
a traffic problem. Traffic state estimation plays an important role in traffic control. Due to the technical and economic limitations, it is difficult to directly measure the system state everywhere thus it requires a traffic observer that can forecasting the system states with partially observed data.
The traffic dynamics is described by the Aw–Rascle–Zhang (ARZ) PDEs model,
\begin{subequations}
\begin{align}
\partial_t \rho + \partial_x(\rho v) & = 0\\
\partial_t v + \left(v-\rho - p'(\rho)\right) \partial_x v & = \frac{V(\rho) - v}{\tau} 
\end{align}
\end{subequations}
where the state variable $\rho(x, t)$ denotes the traffic density and $v(x, t)$ denotes the traffic speed. $V(\rho)$ denotes the equilibrium speed-density relationship, a decreasing function of density. 

The backstepping observer is designed in \cite{yu2020pde},
\begin{subequations}\label{eq:ARZ_obs}
\begin{align}
\partial_t \hat{\rho} + \partial_x (\hat{\rho} \hat{v}) & = \frac{1}{v^*} \left(\exp(-\frac{L_r}{\tau \lambda_1})E_w - E_v\right)\\
\partial_t \hat{v} + (\hat{v} + \hat{\rho} V'(\hat{\rho})) \partial_x \hat{v} & = \frac{V(\hat{\rho}) - \hat{v}}{\tau} + \frac{\lambda_1 - \lambda_2}{q^*} E_v
\end{align}
\end{subequations}
where $L_r$ is the freeway length, with boundary conditions
\begin{align}
\hat{\rho}(0, t) = \frac{y_q(t)}{\hat{v}(0, t)}\,, \quad
\hat{v}(L_r, t) = y_v(t)\,,
\end{align}
and the boundary measurement of the incoming traffic flow  $y_q(t) = q(0, t) = \rho(0, t) v(0, t)$,   outgoing velocity  $y_v(t) = v(L_r, t)$, and outgoing traffic flow $y_{out}(t) = q(L_r, t) = \rho(L_r, t) v(L_r, t)$, which  appears in the output error injection terms $E_{w}$ and $E_v$ in \eqref{eq:ARZ_obs}. \cite[Theorem 3]{yu2020pde} shows the observer design~\eqref{eq:ARZ_obs} guarantees exponential convergence. 
The numerical solution of the nonlinear ARZ PDEs and the nonlinear boundary observers are solved with the Lax–Wendroff method. For collecting the training and test data, we sample the reference density $\rho^{*}$ uniformly between $[110, 130]$ vehicles/km, and reference velocity $v^{*}$ uniformly between [10, 12] m/s. Other parameters are the same as Table 1 in \cite{yu2020pde}.

Performance comparison between different ML-accelerated observers are provided in Table \ref{table_example3}. 
We also compare the computational efficiency of the backstepping PDE observers in~\cite{yu2020pde}, which takes 0.839s to solve each instance, whereas the ML-accelerated method only takes 0.0051s for each instance, which achieves a hundredfold acceleration while keeping the high estimation accuracy.
\begin{table}[t]
\caption{Experiment Results on the PDE Observer for Highway Traffic Density and Velocity Estimation}
\label{table_example3}
\begin{center}
\begin{tabular}{l|l |l | l}
\hline
\bf{Model} & \makecell{\bf{Feedforward} \\ \bf{Neural Observer}}  & \makecell{\bf{Recurrent} \\ \bf{Neural Observer}} & \bf{LSTM}\\
\hline
Parameters & 2,368,290 & 550,018 & 6,634,202\\
\makecell{Training \\ Time (Hours)}   & 0.53 & 3.23 & 0.051\\
Training Error & 5.23e-3 & 2.61e-2 &1.04e-2\\  
Test Error &  5.51e-3 & 9.42e-2 & 1.78e-2\\ 
\hline
\end{tabular}
\end{center}
\end{table}

\section{CONCLUSIONS}
We proposed an ML-accelerated PDE observer design framework which leverages the PDE backstepping design of observer gains and FNO for model learning. It highlights the benefits of inheriting the mathematical rigor of the PDE backstepping observers and obtain significant computational acceleration. We demonstrate the performance of the ML-accelerated observers in three benchmark PDE examples. There are many interesting open questions that remain. 
One important direction is to combine the ML-accelerated observers with downstream control tasks and study their real-time performance, especially in the presence of unmodeled PDE system disturbances.






\vspace{-6pt}
\bibliographystyle{IEEEtran}
\bibliography{reference}

\end{document}